\documentclass[runningheads]{llncs}
\usepackage[T1]{fontenc}
\usepackage{tgbonum}
\usepackage{tgcursor}
\usepackage{hyperref}
\usepackage{booktabs}
\usepackage{tabularx}
\usepackage{threeparttable}
\usepackage{pdflscape} 
\usepackage{adjustbox}
\usepackage{graphicx}
\usepackage{tikz,lipsum,lmodern}
\usepackage{array}
\usepackage{tabularx}
\usepackage[many]{tcolorbox} 
\newtcolorbox{boxK}{
    sharpish corners, 
    boxrule = 0pt,
    toprule = 4.5pt, 
    enhanced,
    fuzzy shadow = {0pt}{-2pt}{-0.5pt}{0.5pt}{black!35} 
}
\begin{document}
\title{Making AI Agents Evaluate Misleading Charts without Nudging}
%
%
\author{Swaroop Panda}
\authorrunning{S Panda}
\institute{Northumbria University}
%
\maketitle              
\begin{abstract}
AI agents are increasingly used as low-cost proxies for early visualization evaluation. In an initial study of deliberately flawed charts, we test whether agents \textit{spontaneously penalise} chart junk and misleading encodings without being prompted to look for errors. Using established scales (BeauVis and PREVis), the agent evaluated visualizations containing decorative clutter, manipulated axes, and distorted proportional cues. The ratings of aesthetic appeal and perceived readability often remained relatively high even when graphical integrity was compromised. These results suggest that un-nudged AI agent evaluation may underweight integrity-related defects unless such checks are explicitly elicited.


\end{abstract}

\section{Introduction}

AI agents for usability testing \cite{lu2025uxagent} represent an emerging area within HCI-AI research \cite{panda2024reflections} . These agents are designed to complement rather than replace traditional user evaluations, serving as a cost-effective preliminary assessment tool \cite{nielsen1992finding} before conducting comprehensive user studies. This approach enables teams to identify potential usability issues efficiently during early development stages.

Shao et al. \cite{shao2025language} find that \textit{``Agents can simulate ratings relatively aligned with human responses.''} This alignment suggests that AI agents can provide meaningful preliminary insights into user experience quality before engaging actual participants. The capability to generate human-like evaluative responses enables design teams to iterate rapidly on interface concepts and identify critical usability concerns early in the development cycle. Furthermore, this preliminary screening can help researchers allocate their limited user testing resources more strategically by focusing human evaluation efforts on the most refined design iterations.

Chart junk \cite{lan2024came} refers to unnecessary decorative elements in visualizations that obscure data comprehension without adding informational value. Beyond mere aesthetic clutter, visualizations can deliberately mislead audiences through manipulated axes, truncated scales, inappropriate chart types, or distorted proportions. These deceptive practices exploit visual perception to misrepresent data relationships, magnify trivial differences, or obscure unfavorable trends. Such manipulations undermine data integrity and can significantly influence decision-making processes based on flawed interpretations.

We investigate how AI agents evaluate visualizations containing chart junk through the application of established assessment scales. Our methodology does not explicitly inform the agents about the presence of decorative or misleading elements. Instead, we prompt agents to respond to scale-based evaluation questions, allowing us to observe whether they organically identify and account for these problematic design features in their assessments. We attempt to infer these organic identifications through systematic analysis of the scale results and response patterns.

\section{Method}
We employ deliberately flawed visualizations in our study because their known deficiencies provide a controlled testing environment for evaluating AI agents' critical assessment capabilities. Importantly, we do not inform the agents that these visualizations contain errors, nor do we explicitly prompt them to identify flaws. Instead, we assess these visualizations through established evaluation frameworks to observe how agents rate them organically. While Panda \cite{panda2025look} deployed a similar methodology, our approach differs in that we do not provide user personas to the AI agents, allowing them to evaluate visualizations independently without predetermined contextual constraints. We utilize two validated evaluation instruments for this assessment: the BeauVis \cite{he2022beauvis} scale and the PREVIS Scale \cite{cabouat2024previs}, which together enable systematic examination of how AI agents perceive and rate visualization quality across multiple dimensions. For the visualisations we randomly pick ten visualisations from the corpus provided by \cite{lan2024came}. We used a single AI agent based on GPT-5.2, accessed via the ChatGPT Plus interface, to evaluate each visualization using the BeauVis and PREVis rating items. Across all the visualizations, the agent was not alerted to potential flaws and provided scale ratings in a single pass per visualization.
\section{Results \& Analysis}

\begin{table}[ht]
\centering
\caption{BeauVis aesthetic ratings of ten visualisations.  Ratings use a seven point Likert scale from 0 (strongly disagree) to 6 (strongly agree).}
\begin{tabular}{l|ccccc}
\hline
\textbf{Visualisation} & \textbf{Enjoyable} & \textbf{Likable} & \textbf{Pleasing} & \textbf{Nice} & \textbf{Appealing} \\
\hline
1 & 3 & 4 & 4 & 4 & 4 \\
2 & 3 & 3 & 3 & 3 & 3 \\
3 & 2 & 2 & 2 & 2 & 2 \\
4 & 4 & 4 & 4 & 4 & 4 \\
5 & 5 & 5 & 5 & 5 & 5 \\
6 & 3 & 3 & 3 & 3 & 3 \\
7 & 2 & 2 & 2 & 2 & 2 \\
8 & 6 & 6 & 6 & 6 & 6 \\
9 & 2 & 2 & 2 & 2 & 2 \\
10 & 3 & 3 & 3 & 3 & 3 \\
\hline
\end{tabular}
\end{table}

\begin{table}[ht]
\centering
\caption{Understand and Layout scales of the PREVis instrument for ten visualisations (ratings 0–7; 0 = strongly disagree, 7 = strongly agree).}
\begin{tabular}{l|ccc|ccc}
\hline
\textbf{Visualisation} & \textbf{obvious} & \textbf{represent} & \textbf{understand} & \textbf{messy} & \textbf{crowd} & \textbf{distract} \\
\hline
1 & 2 & 2 & 2 & 1 & 1 & 2 \\
2 & 6 & 6 & 6 & 6 & 6 & 6 \\
3 & 4 & 4 & 4 & 5 & 4 & 4 \\
4 & 6 & 6 & 6 & 6 & 6 & 6 \\
5 & 6 & 6 & 6 & 6 & 6 & 6 \\
6 & 5 & 5 & 5 & 6 & 6 & 5 \\
7 & 1 & 1 & 1 & 0 & 0 & 0 \\
8 & 5 & 5 & 5 & 6 & 5 & 6 \\
9 & 3 & 3 & 3 & 2 & 2 & 2 \\
10 & 5 & 5 & 5 & 6 & 6 & 6 \\
\hline
\end{tabular}
\end{table}

\begin{table}[ht]
\centering
\caption{Reading data and Reading features scales of the PREVis instrument (ratings 0–7; 0 = strongly disagree, 7 = strongly agree).}
\begin{tabular}{l|ccc|cc}
\hline
\textbf{Visualisation} & \textbf{find} & \textbf{identify} & \textbf{information} & \textbf{visible} & \textbf{see} \\
\hline
1  & 2 & 2 & 2 & 0 & 0 \\
2  & 7 & 6 & 6 & 5 & 5 \\
3  & 4 & 4 & 4 & 3 & 3 \\
4  & 7 & 6 & 6 & 6 & 6 \\
5  & 6 & 6 & 6 & 4 & 4 \\
6  & 5 & 5 & 5 & 3 & 3 \\
7  & 0 & 1 & 1 & 0 & 0 \\
8  & 5 & 5 & 5 & 4 & 4 \\
9  & 2 & 2 & 2 & 2 & 2 \\
10 & 6 & 6 & 6 & 5 & 5 \\
\hline
\end{tabular}
\end{table}


On the BeauVis aesthetic ratings, the ten visualisations show a clear tiering. Visualisation 8 is a ceiling case (all items = 6), followed by Visualisation 5 (all = 5) and Visualisation 4 (all = 4), whereas Visualisations 3, 7, and 9 cluster at uniformly low ratings (all = 2). Across the dataset, responses are highly consistent across the five BeauVis descriptors, suggesting that the agent treated them as a largely unitary judgement of aesthetic appeal rather than as distinct dimensions.

On PREVis (perceived readability), Visualisation 4 is the most consistently readable, with Visualisations 2, 5, and 10 forming a strong second tier across the subscales. Visualisation 5 is particularly informative: despite very high scores on Understand, Layout, and Reading Data, it is comparatively lower on Reading Features (mean = 4.0), indicating that extracting values may be easier than perceiving higher-level visual patterns (e.g., trends or salient deviations). Visualisations 6 and 8 show a similar, though less pronounced, “values > features” profile, while Visualisations 1 and especially 7 underperform across subscales, indicating difficulties both in interpreting the encoding (Understand) and in extracting information (Reading Data/Reading Features). Finally, the PREVis Layout items are typically framed in terms of the absence of clutter (e.g., not messy, not crowded, not distracting); accordingly, higher layout ratings should be interpreted as clearer layouts rather than greater messiness.

\section{Discussion}
Although all ten visualizations contained known design flaws - ranging from chart-junk embellishments to integrity-relevant distortions (e.g., problematic encodings, axis treatments, or rhetorical framing) - the ratings illustrate how apparent quality can decouple from graphical integrity. BeauVis is explicitly an aesthetic instrument rather than a correctness check, and the clear tiering (e.g., Visualisation 8 > 5 > 4, with 3/7/9 low) shows that the agent produced coherent aesthetic judgements across the five descriptors even when the underlying visualization was flawed. PREVis likewise operationalises perceived readability; consequently, strong PREVis profiles (e.g., Visualisations 4/2/10) are entirely compatible with the possibility that a chart can feel legible and well-structured while still being misleading in what it implies. In other words, these results are consistent with the core concern signalled by the title: a visualization may “look right” on standard preference/readability measures while remaining “wrong” in integrity terms.

Substantively, the juxtaposition is most informative where “good presentation” and “flawed content” plausibly coincide. Visualisations that score poorly on both BeauVis and PREVis (e.g., 7) are unlikely to be persuasive, because they fail both aesthetically and in perceived readability. By contrast, a visualization that is rated as highly readable (and at least moderately appealing), such as 4, is a more plausible candidate for unwarranted trust: it exhibits the surface signals that evaluators may use as heuristics for credibility when explicit integrity checks are absent. Notably, the recurring “values > features” pattern in PREVis (e.g., 5, and to a lesser extent 6 and 8) further suggests that even when an agent can extract values or locate information, it may be less sensitive to higher-level visual patterns that often carry the rhetorical thrust of misleading graphics (e.g., exaggerated trends, selective emphasis, or visually amplified differences). That asymmetry is consequential because many integrity violations operate at the level of impression formation rather than single-value lookup.

One implication is that BeauVis and PREVis should be treated as measuring what they claim to measure - affect and perceived readability - rather than as proxies for validity. If AI agents are to be used as early-stage “screeners,” their role may be best conceived as complementary: they can identify charts that are likely to be experienced as pleasing or legible, but those same charts may require additional scrutiny precisely because they appear trustworthy. This motivates a straightforward design recommendation for both research and practice: pair preference/readability instruments with explicit graphical-integrity checks (e.g., axis and proportionality checks, distortion metrics or structured prompts that explicitly ask the agent to assess whether the encoding preserves quantitative relationships). 

\bibliographystyle{splncs04}
\bibliography{paper}

@article{lu2025uxagent,
  title={UXAgent: A System for Simulating Usability Testing of Web Design with LLM Agents},
  author={Lu, Yuxuan and Yao, Bingsheng and Gu, Hansu and Huang, Jing and Wang, Jessie and Li, Yang and Gesi, Jiri and He, Qi and Li, Toby Jia-Jun and Wang, Dakuo},
  journal={arXiv preprint arXiv:2504.09407},
  year={2025}
}

@article{panda2024reflections,
  title={Reflections on emerging HCI--AI research},
  author={Panda, Swaroop and Roy, Shatarupa Thakurta},
  journal={AI \& SOCIETY},
  volume={39},
  number={1},
  pages={407--409},
  year={2024},
  publisher={Springer}
}

@article{panda2025look,
  title={Look into your Heart--Prototypes for a Speculative Design Exploration of Personal Heart Rate Visualization},
  author={Panda, Swaroop},
  journal={arXiv preprint arXiv:2511.07600},
  year={2025}
}

@inproceedings{nielsen1992finding,
  title={Finding usability problems through heuristic evaluation},
  author={Nielsen, Jakob},
  booktitle={Proceedings of the SIGCHI conference on Human factors in computing systems},
  pages={373--380},
  year={1992}
}

@article{shao2025language,
  title={Do language model agents align with humans in rating visualizations? an empirical study},
  author={Shao, Zekai and Shan, Yi and He, Yixuan and Yao, Yuxuan and Wang, Junhong and Zhang, Xiaolong and Zhang, Yu and Chen, Siming},
  journal={IEEE Computer Graphics and Applications},
  year={2025},
  publisher={IEEE}
}

@article{he2022beauvis,
  title={BeauVis: A validated scale for measuring the aesthetic pleasure of visual representations},
  author={He, Tingying and Isenberg, Petra and Dachselt, Raimund and Isenberg, Tobias},
  journal={IEEE Transactions on Visualization and Computer Graphics},
  volume={29},
  number={1},
  pages={363--373},
  year={2022},
  publisher={IEEE}
}

@article{cabouat2024previs,
  title={PREVis: Perceived readability evaluation for visualizations},
  author={Cabouat, Anne-Flore and He, Tingying and Isenberg, Petra and Isenberg, Tobias},
  journal={IEEE Transactions on Visualization and Computer Graphics},
  year={2024},
  publisher={IEEE}
}

@article{lan2024came,
  title={“I Came Across a Junk”: Understanding Design Flaws of Data Visualization from the Public's Perspective},
  author={Lan, Xingyu and Liu, Yu},
  journal={IEEE Transactions on Visualization and Computer Graphics},
  year={2024},
  publisher={IEEE}
}

\appendix
\section{Appendix}
List of the 10  visualizations used in the study.\\
\texttt{tumblr\_mucub3MIiP1sgh0voo1\_1280, tumblr\_n7u9tu3Qes1sgh0voo1\_640,\\ tumblr\_n72ngdGLoY1sgh0voo1\_1280, tumblr\_na7tfshSc41sgh0voo1\_1280,\\
tumblr\_nf7po2WQPT1sgh0voo1\_1280,
tumblr\_ns9cjm7tEy1sgh0voo1\_1280,\\
tumblr\_o5u3u4XMlw1sgh0voo1\_640,
tumblr\_o56n3i2YPW1sgh0voo1\_1280\\
tumblr\_ohq9muiS0C1sgh0voo1\_1280,
tumblr\_onh82iERVC1sgh0voo1\_640}

%




\end{document}